# Using MOEAs to Outperform Stock Benchmarks in the Presence of Typical Investment Constraints

Andrew Clark[1] and Jeff Kenyon[2], *Thomson Reuters*

**Abstract**

Portfolio managers are typically constrained by turnover limits, minimum and maximum stock positions, cardinality, a target market capitalization and sometimes the need to hew to a style (such as growth or value). In addition, portfolio managers often use multifactor stock models to choose stocks based upon their respective fundamental data.

We use multiobjective evolutionary algorithms (MOEAs) to satisfy the above real-world constraints. The portfolios generated consistently outperform typical performance benchmarks and have statistically significant asset selection.

In finance, a portfolio is a collection of assets held by an institution or a private individual. The portfolio selection problem seeks the optimal way to distribute a given monetary budget on a set of available assets. The problem usually has two criteria: expected return to be maximized and risk to be minimized. Classical mean-variance portfolio selection aims at simultaneously maximizing the expected return of the portfolio and minimizing portfolio risk. In the case of linear equality and inequality constraints, the problem can be solved efficiently by quadratic programming, i.e., variants of Markowitz's critical line algorithm. What complicates this simple statement of portfolio construction are the typical real-world constraints that are by definition non-convex, e.g. cardinality constraints which limits the number of assets in a portfolio and minimum and maximum buy-in thresholds. In what follows, we use multi-objective evolutionary algorithms (MOEAs)[3] as an active set algorithm optimized for portfolio selection. The MOEAs generate the set of all feasible portfolios (those portfolios meeting the constraints), calculates the efficient frontier for each and also their respective Sharpe ratio. The portfolio with the best Sharpe ratio becomes the portfolio used for the next time period. We chose MOEAs to solve a non convex optimization problem because there are certain outstanding problems in terms of their use: 1) In the literature MOEAs have not been used to solve multi-period financial problems (or multi-period problems in general), 2) The number and types of constraints in a real world financial portfolio problem exceeds what has been done with MOEAs so far and 3) It is not known if MOEA stock selection is statistically significant. We answer all of these questions with a yes thereby advancing the understanding and use of MOEAs. This especially true when it comes to solving a moderately difficult, multi-period real world problem such as those encountered in finance.

---

[1] Andrew Clark, Chief Index Strategist, Thomson Reuters Indices & Lipper, andrew.clark@thomsonreuters.com
[2] Jeff Kenyon, Lead Software Engineer, Thomson Reuters Indices, jeff.kenyon@thomsonreuters.com

A multi-objective optimization problem (MOP) differs from a single objective optimization problem because it contains several objectives that require optimization. When optimizing a single objective problem, the best single design solution is the goal. But for multiobjective problem with several (possibly conflicting) objectives, there is usually no single optimal solution. Because of this the decision maker is required to select a solution from a finite set of possible solutions by making compromises. A suitable solution should provide acceptable performance over all objectives. The main motivation for using evolutionary algorithms (EAs) to solve multi-objective optimization problems is that EAs can deal simultaneously with a set of possible solutions which allows us to find several members of what is called the Pareto optimal set[4] in a single run of the algorithm. This differs from deterministic mathematical programming techniques where a series of separate runs is required. Additionally EAs are less susceptible to the shape or continuity of the Pareto front, e.g., they can easily deal with discontinuous and concave Pareto fronts. Discontinuity and concavity problems are known obstacles for deterministic mathematical programming.

---

[3] See Coello et al *Evolutionary Algorithms for Solving Multi-Objective Problems*, Kluwer, 2002, for a good introduction to MOEAs.

Adapting any stochastic optimization algorithm (such as an EA) so it can perform a multiobjective optimization requires a change to the method of archiving possible solutions. Any solution on the Pareto front can be identified formally by the fact that it is not *dominated* by any other possible solution. A solution *X* is said to be dominated by solution *Y* if *Y* is at least as good on all counts (constraints) and better on at least one constraint. Stated mathematically:

$$f_i(Y) \leq f_i(X) \forall i = 1, M \text{ and } f_i(Y) < f_i(X) \text{ for some } i$$

As several possible solutions can be generated, an archive of the non-dominated (Pareto optimal) solutions needs to be maintained. A possible archiving scheme is:

- All feasible solutions (Pareto optimal vectors) generated are candidates for archiving

---

[4] A solution or a set is considered Pareto optimal if there exists no feasible solution which would decrease some constraint without causing a simultaneous increase in at least one other constraint.

- If a candidate solution dominates any existing members of the archive, the dominated solutions are removed
- If the new solution is dominated by any existing member of the archive, the new solution is not archived
- If the new solution neither dominates nor is dominated by any members of the archive, the new solution is added to the archive

Using such a scheme, as the search progresses, the archive will converge to the true trade-off surface between constraints.

As to the EA part of MOEA, a generic EA assumes a discrete search space $H$ and a function

$$f : H \rightarrow \mathbb{R}$$

where $H$ is a subset of the Euclidean space $\Re$ (in a multiobjective problem $H$ is a subset of the Euclidean space $\Re^M$ where $M$ is the number of constraints).

The general problem is to find

$$\arg\min_{X \in H} f$$

where $X$ is a vector of the decision variables and $f$ is the objective function.

With EAs it is customary to distinguish *genotype* – the encoded representation of the variables from *phenotype* – the set of variables themselves. The vector $X$ is represented by a string (or chromosome) $s$ of length $l$ made up of symbols drawn from an alphabet $A$ using the mapping

$$c : A^l \rightarrow H$$

If the domain of $c$ is total, i.e. the domain of $c$ is all of $A^l$, $c$ is called a decoding function. The mapping $c$ is not necessarily surjective. The range of $c$ determines the subset of $A^l$ available for exploration by an evolutionary algorithm.

The range of $c$, $\Xi$

$$\Xi \subseteq A^l$$

is needed in order to account for the fact that some strings in the image $A^l$ under $c$ may represent invalid solutions to the original problem.

The search space $\Xi$ can be determined by either Shannon or $2^{nd}$ order Renyi entropy. If the decision variables $X$ are independent, Shannon entropy applies. If the decision variables are correlated then $2^{nd}$ order Renyi entropy applies. A minimization of either entropy will help define the feasible search space $\Xi$.

The string length $l$ depends on the dimensions of both $H$ and $A$ with the elements of the string corresponding to *genes* and values to *alleles*. This statement of genes and alleles is often referred to as *genotype-phenotype mapping*.

Given the statements above, the optimization becomes:

$$\arg\min_{S \in L} g$$

given the function

$$g(s) = f(c(s))$$

Finally, in EAs it is helpful if $c$ is a bijection. The important property of a bijection as it applies to EAs is that bijections have an inverse, i.e. there is unique vector $x$ for every string and a unique string for each $x$.

EAs involve techniques that implement mechanisms inspired by evolution such as reproduction, mutation, recombination, selection and survival of the fittest. Candidate solutions to the optimization problem play the role of individuals in a population and the objective function determines the environment within which the solutions "live." Evolution of the population then takes place after the repeated application of the above operators[5].

In this process, there are two main forces that form the basis of EAs: recombination and mutation which create the necessary diversity and thereby facilitate novelty. Selection acts as a force increasing quality.

Many aspects of EAs are stochastic. Changed pieces of information due to recombination and mutation are randomly chosen. On the other hand, selection operators can be either deterministic, or stochastic. In the latter case, individuals with a higher fitness have a higher chance to be selected than individuals with a lower fitness but typically even weak individuals have a chance to become a parent or to survive.

Choosing the numerical values or techniques that will compute/simulate mutation, recombination and so forth tends to be heuristic. In the MOEA descriptions below, we use standard procedures and values to set the evolutionary parameters. For those interested in reading more about the various values and techniques that are used to select evolutionary parameters, the authors suggest

---

[5] The mathematical development of EA operators and functions is contained in the end note of this paper.

K. Deb, *Multi-Objective Optimization using Evolutionary Algorithms*, John Wiley and Sons, 2000.

The two optimization problems we face are: generate a series of monthly portfolios that outperform the S&P 500 over the last 30 years and generate a set of monthly portfolios that outperform the Russell 1000 Growth index over the last 15 years.

The constraints we will operate under to attain our goals are: turnover is not to exceed 8% per month, the minimum stock position is set at 0.35% of the net asset value of the portfolio, the maximum stock position is set at 4% of the net asset value of the portfolio[6] and a target market capitalization constraint where the average market capitalization of the portfolio must be greater than the average market capitalization of all stocks available to purchase in the current month (this last constraint will mean both portfolio sets will be what is called "large-cap" The S&P 500 and the Russell 1000 Growth are large-cap benchmarks).  Another constraint common to both problems is we must choose stocks that maximize the scores generated by a multi-factor stock model[7]. This constraint typifies the use of what is called fundamental financial data to select stocks that are potential candidates for the final monthly portfolios.

An additional constraint was added just for the Russell 1000 Growth problem: we cannot exceed the average book-to-price value of all stocks available for purchase in the current month.

---

[6] The minimum and maximum position constraints also imply a cardinality constraint. Dividing 100% by 0.35% yields 286 which is the maximum number of stocks any portfolio can have. Dividing 100% by 4% gives 25 which the minimum number of stocks any portfolio can have. The cardinality constraint is solved in the first MOEA while the position constraints are solved in the second MOEA.

[7] The details and back testing of the multi-factor stock model can be found in Chapter 3 of *Handbook of Portfolio Construction*, edited by John Guerard. We thank John for supplying the multi-factor scores and all the other data used in these models.

Meeting this constraint will mean we will generate the required growth portfolios for the Russell 1000 Growth pool.

We solve the issues of the constraints by breaking them into two sets and use two MOEAs. The first MOEA generates potential portfolios that lie within the bounds of all the constraints except turnover and position. In the second MOEA, we trade off the turnover and position constraints as well as mean return and variance (the last being the typical factors used in mean-variance optimization). We set the rebalance period to quarterly versus monthly but stay within the stated turnover constraint (not to exceed 8% per month). The steps each MOEA takes are as follows.

Data Loading, Retrieving and Filtering

1. Load the target candidate constituent sets from the multi-factor score files. "Sets" and "Files" because we need to build two sets of candidate portfolios each month, one for the large-cap portfolio and another for the large-cap growth portfolio (please note that the size of these files goes from approximately 1000 stocks in 1980 to more than 3000 in 2009).
2. Remove candidates that may be excluded, on an *a priori* basis, of being unable to contribute toward the portfolio goals.
   a. Remove candidates with scores of less than 20 (this was a heuristic choice)
   b. Rank candidates by market cap and eliminate the bottom 12% of candidates. If the market cap at this level is greater than US$750M, use US$750M as the floor for the cutoff (this cutoff is based upon common definitions of where small-cap stocks start to appear the U.S. stock market).

3. From this subset of candidate equities, retrieve daily price data going back 287 observations (for use in trading off mean and variance in second MOEA), and going forward 63 observations (for use in performance calculation of final [best] portfolio over the next three months).
4. Calculate the average market cap and average book to price score of the constituents in the target benchmark index (average book-to-price is used to determine if a portfolio satisfies the style constraint).
5. The candidate constituents remaining serve as the primary input to the MOEA phase.

MOEA Phase I

In this phase the goal is to identify a set of portfolios (ideally 50 or less) that will each be examined by the second MOEA. Output of the MOEA for each portfolio is the identification of a subset of the candidates (between 25 and 286) to be used as candidates for later optimization (again 25 because of the 4% maximum constraint and 286 because of 0.35% constraint).

1. The MOEA is invoked, passing the candidate constituent data, portfolio constituents from previous rebalances, the market cap average for the target benchmark index and population, generations, and mutation rate. For the large-cap growth portfolio, there is an additional parameter for the average book-price score of the target benchmark portfolio.
   a. MOEA algorithm is NSGA II, using single point crossover, bit flip mutation, and binary tournament for selection. For the passed parameters, *population* represents the number of proposed solutions that will be carried from generation to generation (note that the number of non-dominated solutions is often considerably smaller); *generations* is the number of generations to be evaluated within the evolutionary algorithm; and *mutation rate* indicates the rate at which a dominated portfolio will be modified as it moves from generation to generation.

   b. For the large-cap portfolio, population is 500, generations are 1200, and mutation rate is 0.03 (3%).

   c. For large-cap growth portfolio, population is 50 (for the three-objective problem, a population of 500 tended to generate several hundred non-dominated solutions; the population number is dropped to avoid generating more solutions than can be explored in the MVO phase). All other parameters are the same.

2. Generation-zero portfolios are started with 156 equities randomly selected. If there are portfolios from a previous rebalance available, those portfolios are seeded into the population.

3. The large-cap portfolio objectives are the maximization of multi-factor score and average market cap. The large-cap growth portfolio adds a third objective to minimize the book-to-price score.

4. Penalties are in place to enforce the cardinality and market cap constraints. For large-cap growth, there is also a penalty for exceeding the book-price average passed in.

5. The MOEA phase produces two files as output. One file contains the objective values for the solution set, and the other contains the portfolios (where each equity has a 0/1 value) describing the Pareto efficient frontier.

6. Any portfolios passing constraints from the previous rebalance are added to the set of portfolios generated by the MOEA

MOEA Phase II

In preparation for running the second MOEA, daily returns are calculated for all equities that may appear in a portfolio (to avoid having to recalculate a daily return series for an equity multiple times), and the daily risk free rate closest to the rebalance date is retrieved (for use in calculating the Sharpe ratio). Then for each portfolio from the previous MOEA, the following actions are taken:

1. Identify the equities designated as candidates for the portfolio being processed, and form the returns matrix. The minimum number of returns to be used is 126 (six months); because the number of observations must exceed the number of candidate equities (the so-called "curse of dimensionality"), the maximum number of returns in the series is determined by portfolio size, up to 287.

2. In addition to forming the returns matrix (step 1 above), we also need to generate mean expected returns, variances, and the covariance matrix.

3. The MOEA is called, using pointers to the above files, a pointer to the file containing the previous winning portfolio, and the MOEA parameters (population = 100, generations = 600, mutation rate = 0.01).

4. Following the lead of an earlier replication of MVO results using MOEA, the algorithm used is SPEA2. SBXCrossover, polynomial mutation, and selection by binary tournament. The SPEA2 archive size is the same as the prior population size.

5. Random portfolios are used for generation zero. If available, the previous winning portfolio is seeded into the generation zero set.

6. In each evaluation, the randomly assigned weights are normalized (so that they sum to 1).

7. In weighting strategy #1 (non-zero values are $0.0035 <= w <= 0.04$), weights greater than 0.00175 are rounded up to 0.0035, while weights under that amount are rounded to zero. Weights over 0.04 are set to 0.04. All weight adjustments are added to a ledger, and then debited or credited at the end of the adjustment process (evenly divided among those that can accept the debit/credit amount without moving outside the set weight boundary).

8. In weighting strategy #2 (all equities are weighted, $0.0035 <= w <= 0.04$), values less than 0.0035 are rounded up to 0.0035, while values over 0.04 are set to 0.04. All weight adjustments are added to a ledger, and then debited or credited at the end of the adjustment process (evenly divided among those that can accept the debit/credit amount without moving outside the set weight boundary).

9. There are four objectives for the MOEA: maximize return, minimize risk, minimize turnover and meet the maximum and minimum holdings constraint. The first two objectives are the standard MVO calculations[8]. The turnover and position objectives calculate the shift in weight between the previous winning portfolio and the proposed portfolio. It attempts to insure that no portfolios break the holdings or turnover bounds.

10. Once all MVOs have been run, the winning portfolio is then selected. Ideally, this is the portfolio with the best Sharpe ratio that has also passed the market cap and turnover constraints. If no portfolios pass the constraints, the portfolio with the best turnover is considered the winner.

11. The performance for the winning portfolio (and the target benchmark index) for the period until the next rebalance is then calculated and logged.

---

[8] See *Investments*, Z. Bodie, A. Kane, A. J. Marcus Richard D. Irwin, Inc., Homewood, IL. 1989, pg. 203, formulas 7.10 and 7.11.

MOEA IIa

If the turnover constraint is not met in Step 9, we use a third MOEA that trades off the best Sharpe ratio portfolio with its existing stock positions against turnover.

The form this MOEA takes is somewhat similar to solving the minimum cut problem in graph theory using MOEAs[9]. The difference between our MOEA and the MOEAs that have been used to solve the minimum cut problem is that we are interested in reducing or enlarging the weight of one or more edges while keeping within the position limit constraints (though we do allow the stock holding to go to zero if needed). Once stocks are re-weighted and the turnover constraint reached, a new efficient frontier is calculated and the new Sharpe portfolio examined. If the new Sharpe portfolio meets the turnover and position constraints, the MOEA in IIa stops and the best Sharpe portfolio for this feasible portfolio from MOEA I is stored. The process is repeated for all the portfolios passed from MOEA I that do not meet the turnover constraints after their first Sharpe portfolio is formed.

Once turnover constraints are met for the Sharpe portfolios that fail step 9 but pass what maybe called Step 9a, Steps 10 and 11 are now executed

In Exhibit 1 are the 1, 3, 5 and 10 year annualized (transaction cost adjusted) returns for the large-cap MOEA portfolios and the S&P 500. The period covered is from December 1979 through December 2009 (121 months)

---

[9] See for example "*Computing Minimum Cuts by Randomized Search Heuristics*," F. Neumann, et al, http://arnetminer.org/viewpub.do?pid=228033

**Exhibit 1**

|  | 1 Year | 3 Year | 5 Year | 10 Year |
|---|---|---|---|---|
| S&P 500 | 9% | 30% | 54% | 137% |
| MOEA | 13% | 44% | 84% | 239% |

Exhibit 2 has the annualized risk and cumulative return on 10,000 USD for the S&P 500 and the MOEA portfolios for the same time period

**Exhibit 2**

|  | Sharpe Ratio | Information Ratio | Cumulative Return on 10,000 USD |
|---|---|---|---|
| S&P 500 | 1.1 | N/A | 37,070 |
| MOEA | 1.8 | 0.14 | 49,500 |

In Exhibit 3 are the 1, 3, 5 and 10 year annualized (transaction cost adjusted) returns for the large-cap growth MOEA portfolios and the Russell Growth 1000. The period covered is from December 1996 through December 2009 (53 months)

**Exhibit 3**

|  | 1 Year | 3 Year | 5 Year | 10 Year |
|---|---|---|---|---|
| R1000 Growth | 10% | 33% | 61% | 159% |
| MOEA | 14% | 48% | 93% | 271% |

Exhibit 4 has the annualized risk and cumulative return on 10,000 USD for the Russell Growth 1000 and the MOEA portfolios for the same time period

**Exhibit 4**

|  | Sharpe Ratio | Information Ratio | Cumulative Return on 10,000 USD |
|---|---|---|---|
| R1000 Growth | 0.14 | N/A | 25,688 |
| MOEA | 0.24 | 0.32 | 31,298 |

As seen in Exhibits 1 – 4 the percent return and USD return of the MOEA portfolios is approximately double the value of their benchmarks.

On a risk-adjusted basis, the results are somewhat mixed. The information ratio for the large-cap growth MOEA is very significant while its Sharpe ratio is only a little larger than the Russell 1000 Growth Sharpe ratio and both are little different from 0 (zero). For the large-cap MOEA, its Sharpe ratio is significantly larger than its benchmark, but its information ratio is very small. So it is not clear that the MOEA portfolios are the better risk-adjusted portfolios in all cases. By not underperforming their benchmarks on a risk-adjusted basis, the implication is that at a minimum the MOEAs reside on a higher curve in risk-return space and could be the more attractive portfolios to investors.

As to the other constraints: 1) In 100% of all cases, the MOEA portfolios on a weighted market capitalization basis met or exceeded the market capitalization constraint, 2) For the smallest and largest positions constraints based on net asset value, none of the MOEA portfolios broke this constraint on either the minimum or maximum side, 3) Turnover did occasionally exceed the 8% limit per month. These occurrences tended to happen early in the 1980's portfolios just as the MOEA was getting on its feet. And the turnover limit was broken in a small number of later portfolios as well. The authors conjecture that in the latter cases the non-dominated feasible solutions handed off to the second MOEA were composed of individual stocks different enough from the prior quarter's portfolio that all the resulting portfolios prevented the turnover constraint from being met. This is an open question however and needs further investigation.

As to one of the reasons why we chose to use MOEAs - is the stock selection of the quarterly portfolios statistically significant - we find the answer to be yes. As measured by John Guerard[10] the MOEA asset selection was very significant. This is a very pleasant surprise to the authors, especially as the portfolios typically contained 150 – 200 stocks. Our results demonstrate that MOEAs can generate statistically significant asset selection while operating under real world constraints and using fundamentally driven stock scores.

In this paper we demonstrate that MOEAs in the presence of real world constraints can generate portfolios that have higher returns than their benchmarks, comparable (if not better) risk adjusted returns versus their benchmarks and statistically significant asset selection.[11]

---

[10] Private communication

[11] We have been asked the question: what are MOEA run times like versus mixed integer programming. We asked a colleague of ours – Todd Morrison – to use the same data supplied to us by John Guerard and the same constraints to solve the same optimization problems but by using mixed integer programming. The *total* time needed to generate *all* the portfolios for each problem was less using MOEAs than mixed integer programming. There is an important caveat however. The major difference in performance occurs before the number of stocks approaches 3000 or more. At that point MOEAs and mixed integer programming take approximately the same time to form the final quarterly portfolio. Further, these test results must be taken with a large grain of salt. The tests discussed here

We arrive at these portfolios by dividing the MOEA in two: the first MOEA generates all the non-denominated feasible sets that meet all the constraints except turnover and minimum and maximum position. The second MOEA trades off the last two constraints along with mean and variance to come up with the final portfolio that has the best Sharpe ratio (or meets the maximum turnover if the other constraints are not met).

As MOEA I can generate feasible solutions to constrained stock selection problems, it could be of help to portfolio managers when deciding which stocks stay in, which move out and which are added to her portfolio. The run time of MOEA I, operating under the constraints above and in the presence of 3000 or more stocks, is 15 – 30 minutes in order to generate a set of 20 - 30 feasible solutions (the run time is less if a smaller solution set is needed). The advantage the feasible solution set gives to the portfolio manager is that at a glance she can see the trade-offs between possible stock portfolios, all of which meet most (if not in some cases all) the constraints to varying degrees.

Finally, as best as we know, this is the first multi-period use of MOEAs in stock portfolio construction. We are encouraged by the results and hope others will extend and improve upon our work.

---

are only two of many problems that can be solved using either technique. A test bed needs to be developed before a clear statement of equality or superiority can be made.

*Acknowledgements*

*The authors would like to thank John Guerard of McKinley Capital for the challenge he set us.*

*We would not have tested MOEAs in the way described above without John's challenge.*

**ENDNOTE**

The 'gory" details of EAs such as the hows and whys of mutation are not described above. What follows are those details.

First we will define the EA fitness function. As $H$ is nonempty set and $c: A^l \to H$ and $f: H \to \mathbb{R}$, we can define the fitness scaling function $T_s: \mathbb{R} \to \mathbb{R}$ and a related fitness function $\Phi \triangleq T_s \circ f \circ c$.

In this definition it is understood that the objective function $f$ is determined by the application, while the specification of the decoding function $c^{12}$ and the fitness scaling function $T_s$ are design issues[13].

Execution of an EA typically begins by randomly sampling with replacement from $A^l$. The resulting collection is the initial population denoted $P$. More generally a population is a collection $P = \{a_1, \ldots, a_\mu\}$ of individuals $a_i \in A^l$. The number of individuals $\mu$ is referred to as the population size.

---

[12] Remember that if the domain of $c$ is total, i.e. the domain of $c$ is all of $A^l$, $c$ is called a decoding function. The mapping $c$ is not necessarily surjective. The range of $c$ determines the subset of $A^l$ available for exploration by the evolutionary algorithm.

Following initialization, execution proceeds iteratively. Each iteration consists of an application of one or more evolutionary operators. The combined effect of the evolutionary operators applied in a particular generation $t \in N$ is to transform the current population $P(t)$ into a new population $P(t+1)$.

In the population transformation, $\mu, \mu' \in \mathbb{Z}^+$ (the parent and offspring population sizes respectively). A mapping $T : H^{\mu} \to H^{\mu'}$ is called a population transformation. If $T(P) = P'$ then $P$ is a parent population and $P'$ is the offspring population. If $\mu = \mu'$ then they are called simply the population size.

The population transform (PT) resulting from an evolutionary operatory (EO) often depends on the outcome of a random experiment. In Merkle and Lamont [14] this result is referred to as a random population transform (RPT) or random PT

To define RPT, let $\mu \in \mathbb{Z}^+$ and $\Omega$ be a set (the sample space). A random function

$R : \Omega \to T(H^{\mu}, \bigcup_{\mu' \in \mathbb{Z}^+} H^{\mu'})$ is called a random population transformation. The distribution of

PTs resulting from the application of an EO depends on the operator parameters, in other words an EO maps its parameters to a RPT.

---

[14] L.D. Merkle and G.B. Lamont, "A Random Function Based Framework for Evolutionary Algorithms", in *Proceedings of the Seventh International Conference on Genetic Algorithms*, Morgan Kauffman, San Mateo, CA., 1997

Now that we have defined both the fitness function and RPT, we can define in general an evolutionary operator: let $\mu \in \mathbb{Z}^+$, $X$ be a set (the parameter space) and $\Omega$ a set. The mapping –

$$Z: X \to T\left(\Omega, T\left[H^\mu, \bigcup_{\mu' \in \mathbb{Z}^+} H^{\mu'}\right]\right)$$

is an evolutionary operator. The set of evolutionary operators is denoted as $EVOP(H, \mu, X, \Omega)$.

There are three common evolutionary operators: recombination, mutation and selection. These three operators are roughly analogous to their similarly named counterparts in genetics. The application of them in EAs strictly follows Darwin - survival of the fittest.

In Merkle and Lamont's definition of the recombination operator $r \in EVOP(H, \mu, X, \Omega)$ If there exists $P \in H^\mu, \Theta \in X$ and $\omega \in \Omega$ such that one individual in the offspring population $r_\Theta(P)$ depends on more than individual of $P$ then $r$ is referred to as a recombination operator.

A mutation is defined in the following manner. Let $m \in EVOP(H, \mu, X, \Omega)$. If for every $P \in H^\mu$, for every $\Theta \in X$ and for every $\omega \in \Omega$ and if each individual in the offspring population $m_\Theta(P)$ depends on at most one individual of $P$ then $m$ is called a mutation operator.

Finally for selection let $s \in EVOP(H, \mu, X \times T(H, \mathbb{R}), \Omega)$. If $P \in H^\mu, \Theta \in X, \Phi: H \to \mathbb{R}$ in all cases and if $s$ satisfies $a \in s_{\Theta,\Phi}(P) \Rightarrow a \in P$ then $s$ is a selection operator.